\documentstyle[epsfig,12pt,aps]{revtex}
\begin{document}
\flushbottom

\widetext
\draft
\title{Coulomb corrections in the calculation of ultrarelativistic heavy ion
production of continuum $e^+ e^-$ pairs}
\author{A. J. Baltz}
\address{
Physics Department,
Brookhaven National Laboratory,
Upton, New York 11973}
\date{\today}
\maketitle

\def\thepage{\arabic{page}}
\makeatletter
\global\@specialpagefalse
\ifnum\c@page=1
\def\@oddhead{Draft\hfill To be submitted to Phys. Rev. A}
\else
\def\@oddhead{\hfill}
\fi
\let\@evenhead\@oddhead
\def\@oddfoot{\reset@font\rm\hfill \thepage \hfill}
\let\@evenfoot\@oddfoot
\makeatother

\begin{abstract}
Coulomb corrections to perturbation theory for producing
electron-positron pairs in ultrarelativistic heavy ion collisions are
considered in a part-analytical, part-numerical approach.  Production
probabilities are reduced from perturbation theory with increasing charge
of the colliding heavy ions, as has been previously argued in the literature.
It is shown here that the reduction from perturbation theory comes from the
appropriate physical spatial cutoff of the electromagnetic potentials arising
from the colliding ultrarelativistic heavy ions.  
\end{abstract}

\makeatletter
\global\@specialpagefalse
\def\@oddhead{\hfill}
\let\@evenhead\@oddhead
\makeatother
\nopagebreak
\section{Introduction}
The problem of calculating heavy ion induced continuum $e^+ e^-$ pair
production to all orders in $Z \alpha$ has received some renewed interest in
the past several years.  Realization that in an appropriate gauge\cite{brbw},
the electromagnetic field of a relativistic heavy ion is to a very good
approximation a delta function in the direction of motion of the heavy ion
times the two dimensional solution of Maxwell's equations in the transverse
direction\cite{ajb1}, led to an exact solution of the appropriate Dirac
equation for excitation of bound-electron positron pairs\cite{ajb2}.  Given
this solution, it was perhaps not surprising that the solution of the Dirac
equation was obtained independently and practically simultaneously by two
different collaborations\cite{sw1,bmc,sw2} for the analagous case of continuum
$e^+ e^-$ pair production induced by the corresponding countermoving delta
function potentials produced by ultrarelativistic heavy ions in a collider
such as RHIC.  An extended discussion and reanalyis of this solution, with
comments on early parallel work in the literature, shortly followed\cite{eg}.
One apparent physical consequence of this solution was that
the rates for pair production in the exact solution agreed with the
corresponding perturbation theory result\cite{bmc,sw2,eg}.

Several authors subsequently argued\cite{serb,lm1,lm2} that a correct
regularization of the exact Dirac equation amplitude should lead to
deviations from perturbation theory, the so called Coulomb corrections. 
Although, as has been pointed out\cite{bg}, the derived exact semiclassical
Dirac amplitude is not simply the exact amplitude for the excitation of a
particular (correlated) electron-positron pair, there are observables,
such as the total pair
production cross section, that can be constructed from this derived amplitude.
The exact amplitude for a correlated electron-positron pair will not be
treated here.  It is the Coulomb corrections to the observables that
{\it can} be constructed from this exact Dirac equation amplitude that are
the topic of this paper.

In what follows it will be shown from a somewhat different approach from what
has been done before that
Coulomb corrections must exist, that they arise from the physical
cutoff of the tranverse Coulomb potential, and the accuracy of their
evaluation has been up to now limited by an effective two-peak approximation
to the exact retarded Dirac amplitude.

\section{The Dirac equation solution}
One begins the semiclassical Dirac solution by representing the
electromagnetic effect of one heavy ion
on the other as the Li\'enhard-Wiechart potential produced by a point
charge on a straight-line trajectory
\begin{equation}
V(\mbox{\boldmath $ \rho$},z,t)={\alpha Z(1-v\alpha_z)\over
\sqrt{ [({\bf b}-\mbox{\boldmath $ \rho$})/\gamma]^2+(z-v t)^2}}
\end{equation}
${\bf b}$ is the impact parameter, perpendicular to the $z$--axis along which
the ions travel, $\mbox{\boldmath $\rho$}$, $z$, and $t$ are the coordinates of
the potential relative to a fixed target (or ion),
$\alpha_z$ is the Dirac matrix,
and $Z, v$ and $\gamma$ are the charge, velocity and relativistic $\gamma$
factor of the moving ion.
If one makes a gauge transformation on the wave function\cite{brbw} 
\begin{equation}
\psi=e^{-i\chi({\bf r},t)} \psi'
\end{equation}
where
\begin{equation}
\chi({\bf r},t)={\alpha Z \over v} 
\ln [\gamma(z-v t)+\sqrt{b^2+\gamma^2(z-v t)^2}]
\end{equation}
the interaction  potential $V({\bf\rho},z,t)$ is gauge transformed to
\begin{equation}
V(\mbox{\boldmath $ \rho$},z,t)={\alpha Z(1-v\alpha_z)\over
\sqrt{ [({\bf b}-\mbox{\boldmath $ \rho$})/\gamma]^2+(z-v t)^2}}
-{\alpha Z(1-(1/v)\alpha_z)\over\sqrt{b^2/\gamma^2+(z-v t)^2}}
\end{equation}

The second term is pure gauge and serves to reduce the range of the
potential in $(z - v t)$ to more
closely map the $(z - v t)$ range of the ${\bf B}$ and ${\bf E}$ fields,
which have the denomenator to
the ${3 \over 2}$ power rather than the ${1 \over 2}$ power of the
untransformed Lorentz gauge potential Eq. (1).

In the ultrarelativistic limit (ignoring correction terms in
$[({\bf b}-\mbox{\boldmath $ \rho$})/\gamma]^2$)\cite{ajb1}
\begin{equation}
V(\mbox{\boldmath $ \rho$},z,t) = - \delta(z - t) (1-\alpha_z) \alpha Z_P
\ln{({\bf b}-\mbox{\boldmath $ \rho$})^2 }.
\end{equation}
This is the potential that allowed the closed form solution of the Dirac
equation for the bound-electron positron problem.  
The full solution of the problem is in perturbation theory form, but with
an eikonalized interaction in the transverse direction
\begin{equation}
V(\mbox{\boldmath $ \rho$},z,t) = - i \delta(z - t) (1-\alpha_z) (\exp[-i
\alpha Z_P \ln{({\bf b}-\mbox{\boldmath $ \rho$})^2 }] - 1 ).
\end{equation}
in place of the pertubation interaction Eq. (5) producing the higher order
effect in $Z \alpha$.  Recall that this
exact semiclassical solution produced a reduction of a little less than 10\% 
in the predicted cross section for Au + Au at RHIC\cite{ajb2}; one can
identify this reduction as a Coulomb correction to bound-electron positron
pair production.

In the bound-electron positron problem one conveniently takes the
electromagnetic field of one moving heavy ion seen in the rest frame of
the heavy ion that receives the created electron.
For production of continuum pairs in an ultrarelativistic heavy ion reaction
one may work in in the center of mass frame and
the electromagnetic interaction goes to the limit of two countermoving
$\delta$ function potentials
\begin{equation}
V(\mbox{\boldmath $ \rho$},z,t)
=\delta(z - t) (1-\alpha_z) \Lambda^-(\mbox{\boldmath $ \rho$})
+\delta(z + t) (1+\alpha_z) \Lambda^+(\mbox{\boldmath $ \rho$}) 
\end{equation}
where
\begin{equation}
\Lambda^{\pm}(\mbox{\boldmath $ \rho$}) = - Z \alpha 
\ln {(\mbox{\boldmath $ \rho$} \pm {\bf b}/2)^2 \over (b/2)^2}.
\end{equation}

The semi-classical Dirac equation with this potential has been solved in closed
form\cite{sw1,bmc,sw2,eg}.
Baltz and McLerran\cite{bmc} noted the apparent agreement of
the obtained amplitude with that of perturbation theory even for large $Z$.
Segev and Wells\cite{sw2} also noted the agreement with perturbation theory
and noted the scaling with $Z_1^2 Z_2^2$ seen in CERN SPS data\cite{vd}.
These data were obtained from reactions of 160 GeV/c Pb ions on C, Al, Pa, and
Au targets as well as 200 Gev/c S ions on the same C, Al, Pa, and Au targets.
The group presenting the CERN data, Vane et al., stated their findings
in summary: ``Cross sections
scale as the product of the squares of the projectile
and target nuclear charges.''  On the other hand, it is well known that
photoproduction of $e^+ e^-$ pairs on a heavy target shows a negative
(Coulomb) correction proportional to $Z^2$ that is well described by the
Bethe-Maximon theory\cite{bem}.

\section{Coulomb corrections}

As noted in the Introduction, several authors have argued that a correct
regularization of the exact
Dirac equation amplitude must lead to Coulomb corrections.  The first
analysis was done in a Weizsacker-Williams approximation\cite{serb}.
Subsequently, Lee and Milstein argued\cite{lm1,lm2} the existence of Coulomb
corrections by an approximate analysis of the closed form solution of the
Dirac equation.  We will take as our starting point
a somewhat extended consideration of the results of Lee and Milstein.

To begin let us write the previously derived semiclassical amplitude for
electron-positron pair production\cite{sw1,bmc,sw2,eg} in the
notation of Lee and Milstein\cite{lm1}
\begin{equation}
M(p,q) = \int { d^2 k \over (2 \pi)^2 } \exp [i\, {\bf k} \cdot {\bf b}]
{\cal M}({\bf k}) F_B({\bf k})
F_A({\bf q_{\perp} + p_{\perp} - k}).
\end{equation}
$p$ and $q$ are the four-momenta of the produced electron and positron
respectively, ${\bf k}$ is an intermediate transverse photon momentum to be
integrated over,
\begin{eqnarray}
{\cal M}({\bf k}) &=& \bar{u}(p) {\mbox{\boldmath $\alpha$}({\bf k -p_{\perp}})
+ \gamma_0 m  \over  -p_+ q_- -({\bf k - p_{\perp}})^2 - m^2 + i \epsilon}
\gamma_- u(-q)
 \nonumber \\
&\ & +  \bar{u}(p) {- \mbox{\boldmath $\alpha$}({\bf k -q_{\perp}})
+ \gamma_0 m  \over  -p_-q_+ -({\bf k - q_{\perp}})^2 - m^2 + i \epsilon}
\gamma_+ u(-q)
\end{eqnarray}
and the effect of the potential Eq. (7-8) is contained in integrals, 
$F_B$ and $F_A$, over the transverse spatial coordinates taking the form
\begin{eqnarray}
F({\bf k}) &=& \int d^2 \rho\, 
\exp [-i\, {\bf k} \cdot \mbox{\boldmath $ \rho$}] 
\{ \exp [-i  2 Z \alpha \ln {\rho}] - 1 \} \nonumber \\
&=& 2 \pi \int_0^{\infty} \rho \, d \rho J_0(k \rho)
\{ \exp [-i  2 Z \alpha \ln {\rho}] - 1 \}.
\end{eqnarray}
$F({\bf k})$ has to be regularized or cut off at large $\rho$.
How it is regularized
is the key to understanding Coulomb corrections.  If one merely regularizes
the integral itself at large $\rho$ one obtains\cite{bmc,sw2,eg} 
apart from a trivial phase
\begin{equation}
F({\bf k}) = {4 \pi \alpha Z \over k^{2 - 2 i \alpha Z} }
\end{equation}

All the higher order $Z \alpha$ effects in $M(p,q)$ are contained only
in the phase of the denominator of Eq. (12).  As we will see, it directly
follows that calculable observables are equal to perturbative results. 

\subsection{Observables}
Before considering the Lee and Milstein analysis, we will discuss the
observables that can be calculated\cite{read,rmgs,mom,obe} from the solution
of a Dirac equation such as Eq. (9-12).  We have pointed out that the derived
semiclassical Dirac amplitude $M(p,q)$ is not simply the exact amplitude for
the excitation of an electron-positron pair\cite{bg}.  The point is that exact
solution of the semi-classical Dirac equation may be used to compute the inclusive
average number of pairs --- not an exclusive amplitude for a particular pair.
Calculating the exact exclusive amplitude to all orders in $Z \alpha$
is not easily tractable due to need for Feynman propagators\cite{bg}.
The possibility of solutions of the semi-classical Dirac equation is
connected to the retarded propagators involved.  In this paper we do not
consider the exclusive (Feynman propagator) amplitude at all.  We concentrate
on observables that {\it can} be construted from the above
amplitude and investigate the Coulomb corrections contained in them.

The occupation number or inclusive number of electrons created in state
$p$ (at impact parameter $b$) is
\begin{equation}
N(p)=\int {m\, d^3 q  \over (2 \pi)^3 \epsilon_q } \vert M(p,q) \vert^2
\end{equation}
Likewise the inclusive number of positrons created in state $q$ is
\begin{equation}
N(q)=\int {m\, d^3 p  \over (2 \pi)^3 \epsilon_p } \vert M(p,q) \vert^2
\end{equation}
These inclusive expressions say nothing about correlations
between electrons in state $p$ and and positrons in state $q$.

The mean number of electron-positron pairs is of
course equal to either the mean number of positrons or the mean number
of electrons and may be obtained by integrating over either of the
previous expressions.
\begin{eqnarray}
N &=& \int {m\, d^3 p  \over (2 \pi)^3 \epsilon_p } N(p) 
= \int {m\, d^3 q  \over (2 \pi)^3 \epsilon_q } N(q) \\
&=& \int {m^2 d^3 p\, d^3 q  \over (2 \pi)^6 \epsilon_p \epsilon_q }
\vert M(p,q) \vert^2 . 
\end{eqnarray}

It is possible to calculate well-defined observables from 
the occupation numbers by integrating over the impact parameter $b$
\begin{equation}
d \sigma(p) = \int d^2  b N(p) = \int d^2 b {m\, d^3 q  \over 
(2 \pi)^3 \epsilon_q } \vert M(p,q) \vert^2,
\end{equation}
\begin{equation}
d \sigma(q) = \int d^2  b N(q) = \int d^2 b {m\, d^3 p  \over 
(2 \pi)^3 \epsilon_p } \vert M(p,q) \vert^2.
\end{equation}
and
\begin{equation}
\sigma_T = \int d^2  b N = \int d^2 b {m^2 d^3 p\, d^3 q  \over (2 \pi)^6
\epsilon_p \epsilon_q } \vert M(p,q) \vert^2.
\end{equation}
$d \sigma(p)$ is the cross section for an electron of momentum $(p)$ where
the state of the positron is unspecified.  Likewise $d \sigma(q)$ is the
cross section for a positron of momentum $(q)$ with
the state of the electron unspecified.
Note that $\sigma_T$ corresponds to a peculiar type of inclusive cross section
which we should call the ``number weighted total cross section'',
\begin{equation}
\sigma_T = \int d^2  b N = \int d^2 b \sum_{n=1}^{\infty} n P_n(b),
\end{equation}
in contrast to the usual definition of an inclusive total cross section
$\sigma_I$ for pair production,
\begin{equation}
\sigma_I = \int d^2 b \sum_{n=1}^{\infty} P_n(b).
\end{equation}

Now we can write for the factor common to all the cross sections
\begin{eqnarray}
\int d^2 b\vert M(p,q) \vert^2 &=& \int d^2 b \int { d^2 k \over (2 \pi)^2 } 
\int { d^2 k' \over (2 \pi)^2 }\exp [i\, {\bf (k - k')} \cdot {\bf b}]
\nonumber \\
&\times& {\cal M}({\bf k}) {\cal M}({\bf k'})^* F_B({\bf k}) F_B({\bf k'})^* 
\nonumber \\
&\times& F_A({\bf q_{\perp} + p_{\perp} - k}) 
F_A({\bf q_{\perp} + p_{\perp} - k'})^* .
\end{eqnarray}
Integrating $\exp [i\, {\bf (k - k')} \cdot {\bf b}]$ over the impact
parameter $b$ in the usual way gives $(2\pi)^2 \delta{\bf (k - k')}$
and so
\begin{equation}
\int d^2 b\vert M(p,q) \vert^2 
= \int { d^2 k \over (2 \pi)^2 } \vert {\cal M}({\bf k}) \vert^2 \vert
F_A({\bf q_{\perp} + p_{\perp} - k}) \vert^2 \vert F_B({\bf k}) \vert^2 . 
\end{equation}

One now obtains expressions for $d \sigma(p)$, $d \sigma(q)$, and $\sigma_T$
that appear identical to the
result of perturbation theory (scaling as $Z_A^2 Z_B^2$) when our previous
expression for $F({\bf k})$ Eq. (12) is employed.
\begin{equation}
d \sigma(p) = \int {m\, d^3 q  \over (2 \pi)^3 \epsilon_q }
\int { d^2 k \over (2 \pi)^2 } \vert {\cal M}({\bf k}) \vert^2
\vert F_A({\bf q_{\perp} + p_{\perp} - k}) \vert^2 \vert F_B({\bf k}) \vert^2 
\end{equation}
\begin{equation}
d \sigma(q) = \int {m\, d^3 p  \over (2 \pi)^3 \epsilon_p }
\int { d^2 k \over (2 \pi)^2 } \vert {\cal M}({\bf k}) \vert^2
\vert F_A({\bf q_{\perp} + p_{\perp} - k}) \vert^2 \vert F_B({\bf k}) \vert^2 
\end{equation}
\begin{equation}
\sigma_T = \int {m^2 d^3 p\, d^3 q  \over (2 \pi)^6
\epsilon_p \epsilon_q }
\int { d^2 k \over (2 \pi)^2 } \vert {\cal M}({\bf k}) \vert^2
\vert F_A({\bf q_{\perp} + p_{\perp} - k}) \vert^2 \vert F_B({\bf k}) \vert^2
\end{equation}
Obviously  $F_B$ and $F_A$ still have to be regularized or cut off at small
$\vert{\bf k}\vert$ and $\vert{\bf q_{\perp} + p_{\perp} - k}\vert$.

\subsection{The regularization of Lee and Milstein}
The strategy of the first paper of Lee and Milstein\cite{lm1} was to evaluate
Coulomb
corrections by Taylor expanding ${\cal M}$ around ${\bf k} =0$, i.e.
${\cal M}({\bf k})\simeq {\bf k \cdot L}$.  The derivative ${\bf L}$
is evaluated at ${\bf k} =0$, and also in the evaluation of e.g. Eq. (26)
${\bf k}$ is ignored in  $F_A({\bf q_{\perp} + p_{\perp} - k})$.
All the ${\bf k}$ dependence of the
integral is then contained in $ d^2 k\, k^2 \vert F_B({\bf k}) \vert^2$.
Lee and Milstein then invite us to consider the integral representing the
difference between the exact solution and the perturbative solution
\begin{equation}
G = \int { d^2 k \over ( 2 \pi )^2 } k^2 [ \vert F({\bf k}) \vert^2
- \vert F^0({\bf k}) \vert^2 ]
\end{equation}
where
\begin{equation}
F({\bf k}) = \int d^2 \rho\, 
\exp [-i\, {\bf k} \cdot \mbox{\boldmath $ \rho$}] 
\{ \exp [-i \chi(\mbox{\boldmath $ \rho$})] - 1 \},
\end{equation}
with the transverse form of the potential not yet specified
\begin{equation}
\chi(\mbox{\boldmath $ \rho$}) = \int_{-\infty}^{\infty} dz
V(z,\mbox{\boldmath $ \rho$}),
\end{equation}
and
\begin{equation}
F^0({\bf k}) = -i\int d^2 \rho\, \exp [-i\, {\bf k} \cdot
\mbox{\boldmath $ \rho$}] \chi(\mbox{\boldmath $ \rho$}) 
\end{equation}
is the perturbative expression limit of $F({\bf k})$.

Lee and Milstein keep the $2 Z \alpha \ln ( \rho)$ form for
$\chi(\mbox{\boldmath $ \rho$})$ but switch the order of integration between
$\rho$ and $k$.  
They integrate $k$ to some finite upper limit
$Q$ an then claim to set $Q$ to infinity in the resulting expression.
Actually $Q$ simply falls out of the problem by a rescaling of
$\rho$ to $\rho / Q$.  Next, after integrating over the rescaled $\rho$, the
expression they obtain is a universal function of $Z \alpha$
\begin{equation}
G = -8 \pi (Z \alpha)^2 [Re \psi ((1 + i Z \alpha) + \gamma_{Euler}],
\end{equation}
where $\psi ((1 + i Z \alpha)$ is the digamma function and $\gamma_{Euler}$ is
Euler's constant.  This expression may be alternatively expressed as
\begin{equation}
G = -8 \pi (Z \alpha)^2 f(Z \alpha),
\end{equation}
where $f(Z \alpha)$ is the same function that was derived by Bethe and Maximon
for Coulomb corrections to $e^+ e^-$ photoproduction on heavy nuclei
and takes the form
\begin{equation}
f(Z \alpha)= (Z \alpha)^2 \sum_{n=1}^\infty {1 \over n(n^2+(Z \alpha)^2)}.
\end{equation}

The derivation and result may seem a little mystifying.  Lee and Milstein
state, ``Thus, we come to a remarkable conclusion: although the main
contribution to the integral in Eq. (4) comes from the region of small $k$,
where $\vert F({\bf k}) \vert$ differs from 
$(\vert F^0({\bf k} \vert) = 4 \pi Z \alpha / k^2$ and depends on the
regularization parameters (the radius of screening), nevertheless the integral
$G$ itself is a universal function of $Z \alpha$.''  As we will see later
the only part of this quoted statement that is completely true is that $G$
is a universal function of $Z \alpha$.

$G$ is then used by Lee and Milstein to calculate the Coulomb correction
arising from ion $B$ by taking ion $A$ is to lowest order in $Z \alpha$.
Generalizing this approach, the corresponding Coulomb correction arising
from ion $A$ is also evaluated\cite{lm2}.  The sum of these two contributions
then agree with the Coulomb corrections as evaluated by Ivanov, Schiller,
and Serbo\cite{serb} using the Weizsacker-Williams method.

\subsection{A physical regularization}
Let us try to understand Lee and Milstein's result by putting in a physical
cutoff to the transverse potential $\chi(\mbox{\boldmath $ \rho$})$ (which has
been up to now set to $2 Z \alpha \ln{\rho}$).  Instead of regularizing the
integral itself and letting the cutoff radius go to infinity
as was originally done\cite{sw1,bmc,sw2,eg}, we will apply an appropriate
physical cutoff to the interaction
potential.  In the Weizsacker-Williams or equivalent photon treatment of
electromagnetic interactions the potential is cut off at impact parameter
$b \simeq \gamma / \omega$, where $\gamma$ is the relativistic boost
of the ion producing the photon and $\omega$ is the energy of the photon.
As Lee and Milstein subsequently recall (but do not
utilize) if
\begin{equation}
\chi(\mbox{\boldmath $ \rho$}) = \int_{-\infty}^\infty dz V(\sqrt{z^2+\rho^2})
\end{equation}
and $V(r)$ is cut off in a physically motivated way, such as an
equivalent photon cutoff, then
\begin{equation}
V(r)={-Z \alpha \exp[-r \omega_{A,B} / \gamma] \over r }
\end{equation}
where 
\begin{equation}
\omega_A= {p_+ + q_+ \over 2 };\  \omega_B={p_- + q_- \over 2 }
\end{equation}
with $\omega_A$ the energy of the photon from ion $A$ moving in the positive
$z$ direction and $\omega_B$
the energy of the photon from ion $B$ moving in the negative $z$
direction.  For simplicity we will suppress the subscripts on $\omega$,
remembering however for future possible use that $\omega_{A,B}$ are well
defined in terms of $p_{\pm}$ and $q_{\pm}$.
The integral Eq. (34) can be carried out to obtain
\begin{equation}
\chi(\rho)= - 2 Z \alpha K_0(\rho \omega / \gamma),
\end{equation}
and
\begin{equation}
F_{A,B}({\bf k}) = 2 \pi \int d \rho \rho J_0(k \rho)
\{\exp [2i Z_{A,B}\alpha K_0(\rho \omega / \gamma)] -1 \} .
\end{equation}
The modified Bessel function
$K_0(\rho \omega / \gamma) = - \ln(\rho\omega / 2 \gamma)$ for small $\rho$
and  cuts off exponentially at $\rho \sim \gamma / \omega $.  This
is the physical cutoff to the transverse potential.

One may define $\xi = k \rho$ and rewrite Eq.(38)
\begin{equation}
F_{A,B}({\bf k}) = {2 \pi \over k^2} \int d \xi \xi J_0(\xi) 
\{\exp [2i Z_{A,B}
\alpha K_0(\xi \omega / \gamma k)] - 1 \}.
\end{equation}
It is now clear that $F_{A,B}$ is a function of
$1/k^2$ times some function of $(\gamma k / \omega)$.
The perturbative limit $F^0_{A,B}({\bf k})$ is analytically soluable and
takes the form 
\begin{equation}
F^0_{A,B}({\bf k}) = {4 \pi Z_{A,B} \alpha \over k^2 + \omega^2 / \gamma^2}
={4 \pi Z_{A,B} \alpha \over k^2 (1 + \omega^2 / k^2 \gamma^2)}
\end{equation}

Fig. (1) displays the results of numerical calculation of the scaled magnitude
of $F({\bf k})$ as a function of 
$k \gamma / \omega$ for $Z = 1$ (essentially the perturbative form Eq. (40))
and for $Z = 82$.  Note that the upper cutoff of $\rho$ at $\gamma / \omega $
has the effect of regularizing $F({\bf k})$ at small $k$.  $F({\bf k})$ goes
to the constant $4 \pi \gamma^2 / \omega^2$ as $k$ goes to zero in the
$Z=1$ perturbative case; it goes to a reduced constant value as $k$ goes to
zero for $Z=82$.  The form of the original solution Eq. (11)
\begin{equation}
F({\bf k}) = {4 \pi \alpha Z \over k^{2 - 2 i \alpha Z} }
\end{equation}
is simply wrong because it is unphysical.  Since it lacks a proper physical
cutoff in $\rho$, it not only blows up at $k = 0$, but it also fails to
exhibit the correct reduction in magnitude that occurs when
$k \gamma / \omega$ is not too large.
\begin{figure}[h]
\begin{center}
\epsfig{file=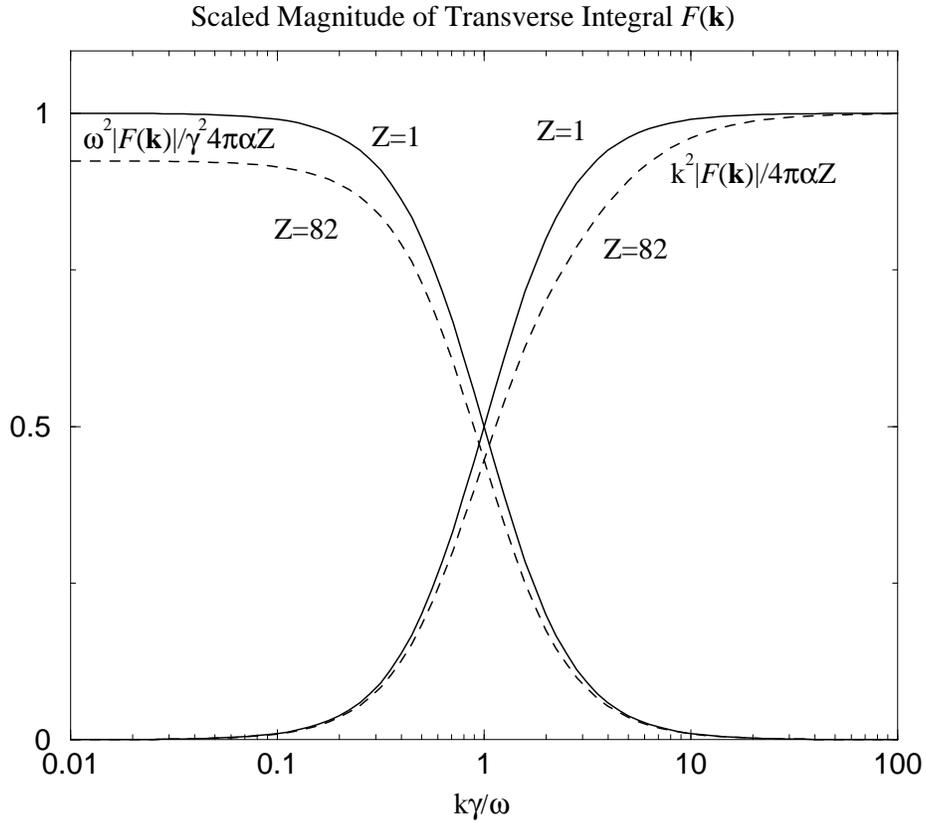,height=12cm}
\end{center}
\caption{The decrease in the magnitude of the
transverse integral $F$ with $Z$.  The two sets of curves have been normalized
to display that the finite Coulomb correction only rescales down the 
$\vert F({\bf k})\vert \sim 1 / \omega^2$ behavior at $k \omega / \gamma = 0$
and that the negative Coulomb corrections do not vanish until well above
the onset of $\vert F({\bf k})\vert \sim 1 / k^2$ dominant behavior.}
\label{graph1_fig}
\end{figure}

Fig. (2) is an alternate display of results of the numerical
calculations showing the fractional decrease in the ratio
$\vert F({\bf k}) \vert / \vert F^0({\bf k})\vert$ for various values of $Z$ 
as a function of
$k \gamma / \omega$.  It is clear from the two Figures that for increasing
$Z$ Coulomb corrections reduce $F({\bf k})$ from the perturbative result for
$k \gamma / \omega << 100$.  Only for $k >\ \sim 100\ \omega / \gamma $ does 
the magnitude of $F({\bf k})$ go over into the original form of Eq. (41).

\begin{figure}[h]
\begin{center}
\epsfig{file=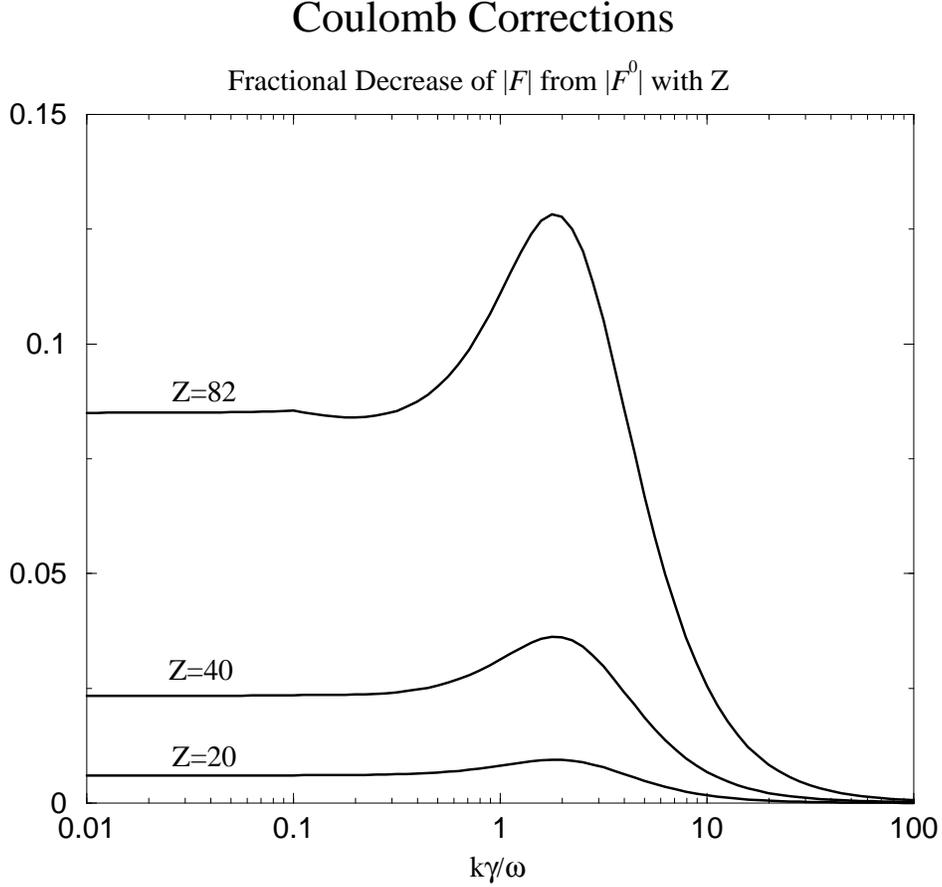,height=12cm}
\end{center}
\caption{The curves display the ratio
$\vert [F_0({\bf k})-F({\bf k})]/F_0({\bf k})$ as a function of Z.}
\label{graph2_fig}
\end{figure}
Now let us consider $G$ again with the specific forms of $F({\bf k})$
displayed in Fig. (1-2)
\begin{equation}
G = \int { d^2 k \over ( 2 \pi )^2 } k^2 [ \vert F({\bf k}) \vert^2
- \vert F^0({\bf k}) \vert^2 ]
\end{equation}
Note that, given the $1 / k^2$ dependence of $F({\bf k})$ and of
Eq. (39-40), this is a logarithmic
integral of $k$ (i.e  $dk/k$) times a function of $k \gamma / \omega$.
Therefore the integration is really over the combination variable
$k \gamma / \omega$.  Thus $\gamma / \omega$ falls out of the integral, and
the Coulomb correction function $G$ does not depend on $\gamma$ or $\omega$.

I have evaluated $G$ numerically and found it exactly converging to Lee and
Milstein's result according to the expected improved precision with
decreasing mesh size.  I attained agreement to one part in $10^6$.

Conjecturing that the detail of the cutoff should not matter, I replaced
the function $K_0(\rho \omega / \gamma)$ with a different function that
also goes as $- \ln(\rho\omega / 2 \gamma)$ (plus an irrelevant constant, 
$1/2 + \gamma_{Euler}$) for small $\rho$ and also cuts off exponentially at
$\rho \sim \gamma/\omega $:
\begin{equation}
L_0({\rho \omega / \gamma) = {(\rho \omega / \gamma)^2 \over 2} 
[K_1^2(\rho \omega / \gamma)
-K_0(\rho \omega / \gamma) K_2(\rho \omega / \gamma)}].
\end{equation}
Calculations of $G$ with $L_0$ in place of $K_0$ similarly converge numerically
to the result of Lee and Milstein with agreement to one part in $10^6$.
Note however the non-identical shapes of the contribution to $G$ as a function
of $ k \gamma / \omega $ for the $K_0$ and $L_0$ transverse potential
forms exhibited in Fig. (3), even though the area above the two curves
(the value of $G$) is identical.   

\begin{figure}[h]
\begin{center}
\epsfig{file=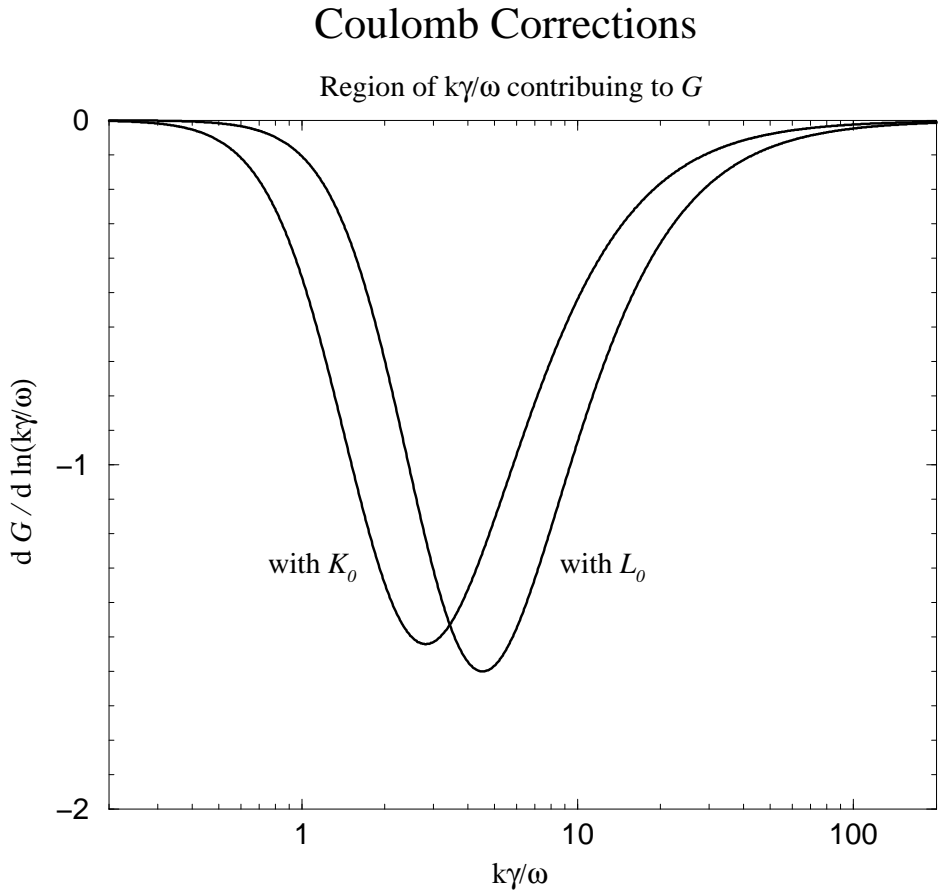,height=12cm}
\end{center}
\caption{Region of $k \gamma / \omega$ contributing to the Coulomb correction
integral $G$ for $Z=82$.}
\label{graph3_fig}
\end{figure}
Now we can begin to understand the result of Lee and Milstein.  The reason
that ``the integral $G$ itself is a universal function of $Z \alpha$'' is
that the first order $k^2$ factor from the expansion makes the integral $G$
logarithmic and so, contrary to what Lee and Milstein state, $G$ does not
``depend...on the regularization parameters (the radius of screening)''.
The radius of screening, i.e. $\gamma / \omega$, is finite, but it has
fallen out of the problem.  Furthermore ``the main contribution to the
integral'' does not ``come from the region of small $k$'' but, as is seen from 
the plot of the physically motivated $K_0$ curve in Fig.(3), the main
contribution is peaked at $ \gamma k / \omega = 2.8 $ and spreads out between
half-maxima at $1.3$ and $7.5$.

Note that the decoupling of the Coulomb corrections from $\gamma / \omega$
seen in $G$ is only valid to first order in $k$.  Including higher order
terms in $k$ or, alternatively carrying out a full numerical evaluation
of e.g. Eq. (26), would necessarily restore some dependence on
$\gamma / \omega$ to the Coulomb corrections.  A previous Monte Carlo
perturbation theory calculation of Bottcher and Strayer\cite{bs} displays
the pair production cross section
as a function of $P_T=p_{\perp}+q_{\perp}$, and  shows a significant
deviation between an exact Monte Carlo evaluation of the cross section
and evaluation using a two peak approximation (in particular see Fig.(9)
of Ref.\cite{bs}).
Since in carrying out their calculation, Lee and Milstein made a variety
of a two peak approximation (assuming $P_T=p_{\perp}+q_{\perp}$ small),
one has to assume that the precision of their results is limited.

\section{General observations}

To lowest order in transverse momentum (small $k$ and small
$P_T=p_{\perp}+q_{\perp}$), Coulomb corrections do exist as a universal
function of $f(\alpha Z)$, where $f(\alpha Z)$ is the same function of
Bethe and Maximon derived for Coulomb corrections to electron-positron
pair photoproduction.  These Coulomb corrections reduce the uncorrelated
electron or positron production cross sections
and the number weighted total pair cross section.

In general and not limited to lowest order in transverse momentum, 
Coulomb corrections are
a function of only $Z$ and the combination variable $k\gamma/\omega$.
Coulomb corrections arise from the finite cutoff of the
transverse spatial integral at $\gamma / \omega$ and vanish for large
$k \gamma / \omega$.

Since the CERN data cover a large part of the momentum range of produced
positrons and scale perturbatively, they still seem to present a puzzle.
It would be useful to carry out full calculations of the total number
weighted cross section $\sigma_T$ as well as of the uncorrelated momentum
dependent electron and positron cross sections $d\sigma(p)$ and $d\sigma(q)$,
utilizing the transverse integrals
with a correct physical cutoff.  Since the CERN data only detects positrons,
comparison with a full calculation of $d\sigma(q)$ is appropriate.

\section{Acknowledgments}
I would like to thank Francois Gelis and Larry McLerran for useful
discussions.  This manuscript has been authored under Contract No.
DE-AC02-98CH10886 with the U. S. Department of Energy. 

\end{document}